\documentclass[showpacs,aps,graphicx,twocolumn]{revtex4}

\usepackage{graphicx}

\begin{document}

\title{Hyperentanglement concentration for two-photon four-qubit systems  with linear optics\footnote{Published in Phys. Rev. A \textbf{88}, 012302 (2013)}}
\author{Bao-Cang Ren, Fang-Fang Du, and Fu-Guo Deng\footnote{Corresponding author: fgdeng@bnu.edu.cn} }
\address{ Department of Physics, Applied Optics Beijing Area Major Laboratory,
Beijing Normal University, Beijing 100875, China}
\date{\today }

\begin{abstract}

Hyperentanglement, defined as the entanglement in several degrees of
freedom (DOFs) of a quantum system, has attracted much attention
recently. Here we investigate the possibility of concentrating the
two-photon four-qubit systems in partially hyperentangled
 states in both the spatial mode and the polarization DOFs with linear optics. We first introduce
our parameter-splitting method to concentrate the systems in the
partially hyperentangled states with  known parameters, including
partially hyperentangled Bell states and cluster states.
Subsequently, we present another two nonlocal hyperentanglement
concentration protocols (hyper-ECPs) for the systems in partially
hyperentangled unknown states, resorting to the Schmidt projection
method. It will be shown that our parameter-splitting method is very
efficient  for the concentration of the quantum systems in partially
entangled states with known parameters, resorting to linear-optical
elements only. All these four hyper-ECPs are feasible with current
technology and they may be useful in long-distance quantum
communication based on hyperentanglement as they require only linear
optical elements.

\end{abstract}

 \maketitle

\section{Introduction}

Entanglement, a unique phenomenon in quantum mechanics theory, has
largely improved the methods of manipulating and transforming
information in quantum information processing and quantum computing
\cite{QC}. The entangled photon systems can act as the quantum
channels in some typical long-distance quantum communication
proposals, such as quantum key distribution \cite{QKD}, quantum
teleportation \cite{QT1}, quantum dense coding \cite{DC1,DC2},
quantum secret sharing \cite{QSS1,QSS2}, and so on. An entangled
photon pair is usually  produced locally, so the environment noise
is inevitable for it in its distribution process among the parties
in quantum communication, which will decrease its entanglement. The
photon signals can only be transmitted no more than several hundreds
of kilometers in an optical fiber or a free space with current
technology, and  quantum repeaters are required to connect the two
neighboring nodes in long-distance quantum communication network. In
the storage of a quantum state in a quantum repeater, the
entanglement of an entangled photon system will also be decreased by
decoherence. The fidelity and the security of long-distance quantum
communication protocols may be decreased with the decrescence of the
entanglement in  photon systems.

Entanglement purification and  entanglement concentration are two
passive ways for depressing the noise effect on entangled systems,
with which some high-fidelity nonlocal entangled systems can be
obtained from a set of less-entangled systems. Entanglement
purification is used to distill a subset of high-fidelity nonlocal
entangled systems from a set of those in a mixed state with less
entanglement
\cite{EPP1,EPP2,EPP3,EPPsimon,EPPexperiment,EPPsheng1,EPPsheng2,
EPPsheng3,EPPdeng1}, while entanglement concentration is used to
distill some nonlocal maximally-entangled systems from a set of
systems in a partially-entangled pure state. In 1996, Bennett
\emph{et al}. \cite{ECP1} introduced the first entanglement
concentration protocol (ECP) for two-photon systems by using the
Schmidt projection method and collective measurements. Since this
pioneering work, many interesting ECPs have been proposed for photon
systems \cite{ECP2,ECP3,ECP4,ECP5,ECP5a,ECP5b,ECP6,ECP7,ECP8,ECP11}.
They can be divided into two groups. In the first group, the ECPs
\cite{ECP1,ECP4,ECP5,ECP6} require that the two remote users, say
Alice and Bob, do not know the parameters about the partially
entangled pure state of the photon systems. In the second group, the
ECPs \cite{ECP2,ECP3,ECP7,ECP8,ECP11} require that Alice and Bob
know the parameters accurately. The latter have a higher efficiency
than that in the former in theory.

Hyperentanglement, defined as the entanglement in several degrees of
freedom (DOFs) of a quantum system, has attracted much attention for
quantum information recently. For example, it has been used to
assist the polarization photonic Bell-state analysis
\cite{BSA1kwiat,BSA2walborn,BSA3,BSA4,BSA}. In 2003, Walborn
\emph{et al}. \cite{BSA} presented a complete Bell-state analysis in
the coincidence basis using hyperentangled states. In 2002, Simon
and Pan \cite{EPPsimon} presented a polarization photonic
entanglement purification protocol (EPP) for a parameter
down-conversion (PDC) source with linear optics, resorting to
spatial-polarization hyperentanglement.  In 2005, Barreiro \emph{et
al}. \cite{GenH} experimentally prepared hyperentangled photon pairs
in polarization, spatial mode, and time energy DOFs with spontaneous
parametric down-conversion photons. In 2008, Sheng \emph{et al}.
\cite{EPPsheng1} presented an efficient EPP for a PDC source with
nonlinear optics, resorting to the spatial-polarization
hyperentanglement. Subsequently, some deterministic EPPs
\cite{EPPsheng2,EPPsheng3,EPPdeng1} were proposed with
hyperentanglement. In 2008, Barreiro \emph{et al}. \cite{HESC} beat
the channel capacity limit of superdense coding with linear optics,
resorting to polarization-orbital-angular-momentum
hyperentanglement. And some hyperentangled Bell-state analysis
protocols \cite{kerr,HBSA,HBSA1,HBSA2} were proposed to increase the
channel capacity of quantum communication recently. In 2007, Wei
\emph{et al}. \cite{HBSA2} divided the 16 hyperentangled Bell states
into seven groups with linear optics. In 2012, an interesting
quantum repeater protocol based on spatial-polarization
hyperentanglement was proposed \cite{wangrepeater}. In 2013, Graham
\emph{et al}. \cite{DCQD} experimentally implemented direct
characterization of quantum dynamics assisted by hyperentanglement.

In this article, we investigate the possibility of concentrating the
two-photon four-qubit systems in the nonlocal partially
hyperentangled states in both the spatial mode and the polarization
DOFs with linear-optical elements. First, we introduce two feasible
nonlocal spatial-polarization hyperentanglement concentration
protocols (hyper-ECPs) for the systems in partially hyperentangled
states with known parameters (including  hyperentangled Bell-class
states and cluster-class states), resorting to our
parameter-splitting method. Subsequently,  we present another two
hyper-ECPs for the partially hyperentangled photon systems in
unknown states, resorting to the Schmidt projection method, a
conventional way for entanglement concentration.  These two
hyper-ECPs can be accomplished with two copies of unknown states and
parity-check measurements. The parity-check measurement on the
polarization DOF is constructed with polarizing beam splitters and
single-photon detectors, and the parity-check measurement on
spatial-mode DOF is constructed with the Hong-Ou-Mandel (HOM) effect
\cite{HOM} of a 50:50 beam splitter and single-photon detectors. It
will be shown that the hyper-ECPs with our parameter-splitting
method are far more efficient  for the concentration of the quantum
systems in partially entangled states with known parameters than
those with the Schmidt projection method.

This paper is organized as follows: In Sec. \ref{sec2}, we present
two hyper-ECPs for  the two-photon four-qubit systems in   known
pure states with our parameter-splitting method, including the one
used for the systems in a known partially hyperentangled Bell state,
which is discussed in Sec. \ref{sec21}, and the other used for those
in a known partially hyperentangled cluster state, discussed in Sec.
\ref{sec22}. In Sec. \ref{sec3}, we discuss the hyperentanglement
concentration for the two-photon four-qubit systems in unknown
partially entangled pure states with linear optics. Some discussions
and a summary are given in Sec. \ref{sec4}. In  Appendix
\ref{appendixa}, a hyper-ECP for  an arbitrary unknown
hyperentangled cluster-class state is discussed. In  Appendix
\ref{appendixb}, the entanglement purification of a mixed
hyperentangled Bell state is given. Our parameter-splitting-based
entanglement concentration for a known partially entangled state in
one DOF is discussed in  Appendix \ref{appendixc}.

\section{Hyper-ECPs  with parameter-splitting method}
\label{sec2}

In this section, we present two hyper-ECPs for two-photon four-qubit
systems in  known nonlocal hyperentangled spatial-polarization pure
states based on our parameter-splitting method with linear-optical
elements. One hyper-ECP is used for two-photon four-qubit systems in
a known partially hyperentangled Bell state and the other is used
for those in a known partially hyperentangled cluster state. In both
these hyper-ECPs, Alice and Bob obtain a subset of two-photon
four-qubit systems in maximally hyperentangled states by splitting
the parameters of the initial nonlocal partially hyperentangled
states with linear-optical elements only.

\begin{figure}[htbp]             
\centering\includegraphics[width=7.82 cm]{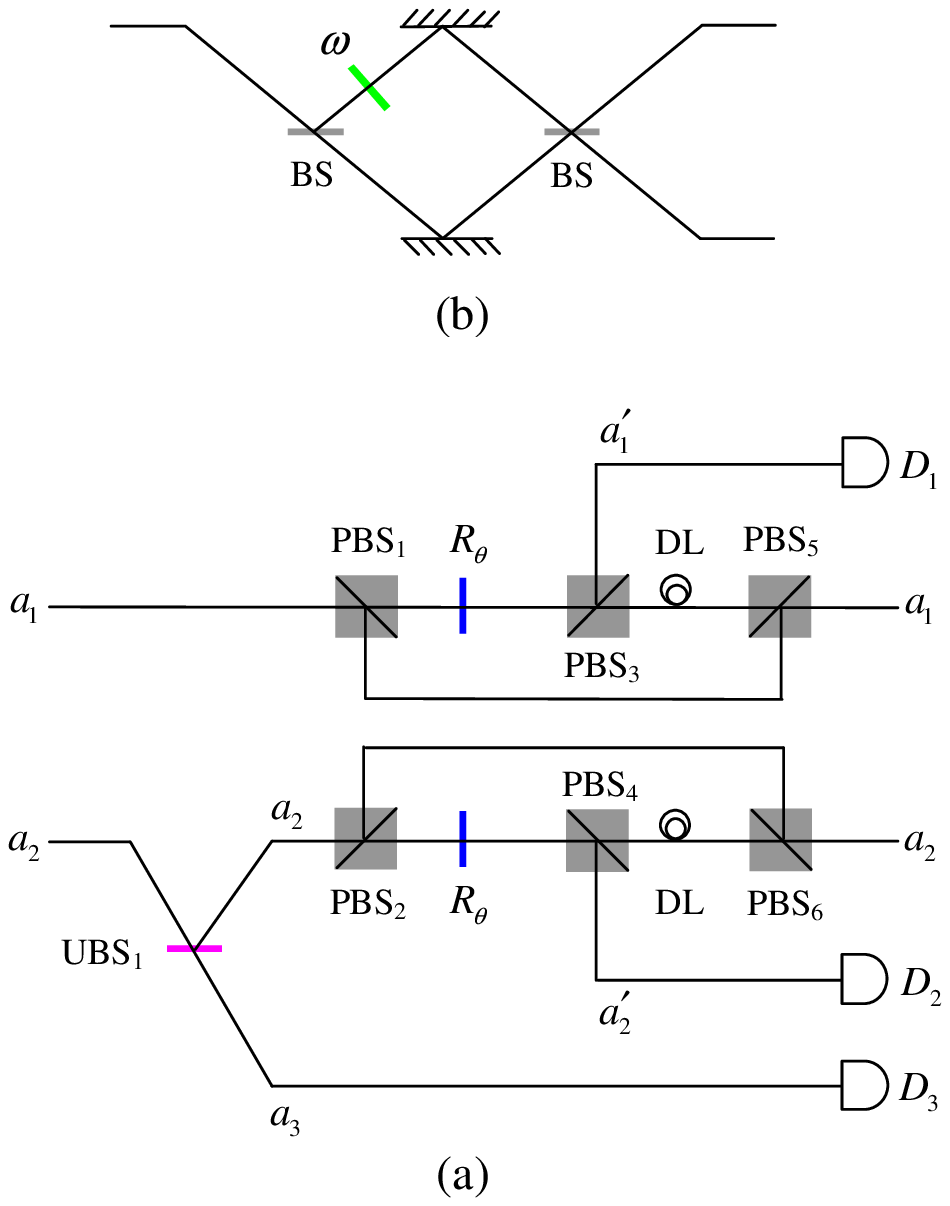} \caption{ (Color
online) (a) Schematic diagram of our hyper-ECP for a partially
hyperentangled Bell state with known parameters. (b) Schematic
diagram of an unbalanced BS (UBS). BS represents a balanced 50:50
beam splitter. $\omega$ represents a phase shift which decides the
reflection coefficient and the transmission coefficient of the
unbalanced BS. $a_1$ and $a_2$ represent the two original spatial
modes of the photon $A$. $a'_1$ and $a'_2$ are another two spatial
modes of the photon $A$. UBS$_1$ represents an unbalanced beam
splitter with the reflection coefficient $R_1=\gamma/\delta$.
$R_\theta$ represents a wave plate which can rotate the horizontal
polarization with an angle $\theta=arccos(\beta/\alpha)$.
PBS$_{\,i}$ ($i=1,2,\cdots, 6$) represents a polarizing beam
splitter, which transmits the photon in the horizontal polarization
$\vert H\rangle$ and reflects the photon in the vertical
polarization $\vert V\rangle$, respectively. DL denotes a time-delay
device which is used to make the two wavepackets reach the last PBS
(PBS$_5$ or PBS$_6$) in each Mach-Zehnder interferometer
simultaneously. $D_1$, $D_2$, and $D_3$ represent three
single-photon detectors.} \label{figure1}
\end{figure}

\subsection{Hyper-ECP for two-photon four-qubit systems in a partially hyperentangled Bell  state}
\label{sec21}

Let us assume that the initial nonlocal hyperentangled Bell-class
state in both the spatial mode and the polarization DOFs  is
\begin{eqnarray}                            
|\varphi_0\rangle_{AB}&=&(\alpha|H\rangle_A|H\rangle_B+\beta|V\rangle_A|V\rangle_B)\nonumber\\
&&\otimes(\gamma|a_1\rangle|b_1\rangle+\delta|a_2\rangle|b_2\rangle).
\end{eqnarray}
Here the subscripts $A$ and $B$ represent the two photons kept by
the two remote users, say Alice and Bob. $|H\rangle$ and $|V\rangle$
represent the horizontal and the vertical polarizations of photons,
respectively. $|a_1\rangle$ ($|b_1\rangle$) and $|a_2\rangle$
($|b_2\rangle$) are the two spatial modes of the photon $A$ ($B$).
$\alpha$, $\beta$, $\gamma$, and $\delta$ are four real parameters
that are known to Alice and Bob, and they satisfy the relation
\begin{eqnarray}                           
|\alpha|^2+|\beta|^2=|\gamma|^2+|\delta|^2=1.
\end{eqnarray}
For describing the principle of our hyper-ECP explicitly and simply,
we assume that $|\alpha| > |\beta|$ and $|\gamma| < |\delta|$ below.
The principle of our hyper-ECP in other cases is the same as this
one with or without a little modification.

The principle of our hyper-ECP for a partially hyperentangled Bell
state is shown in Fig. \ref{figure1}(a).  It can be implemented with
some local unitary operations on the photon $A$ in both the
spatial-mode and the polarization DOFs performed by Alice. Bob
performs no operations in this hyper-ECP. In detail, first, Alice
performs a unitary operation on the spatial mode $a_2$ by using an
unbalanced BS (i.e., UBS$_1$) \cite{UBS} with the reflection
coefficient $R_1=\gamma/\delta$, shown in Fig. \ref{figure1}(b), and
the partially hyperentangled Bell-class state
$|\varphi_0\rangle_{AB}$ is changed to be $|\varphi_1\rangle_{AB}$.
Here
\begin{eqnarray}                             
|\varphi_1\rangle_{AB}&=&(\alpha|H\rangle_A|H\rangle_B+\beta|V\rangle_A|V\rangle_B)\otimes[\gamma(|a_1\rangle|b_1\rangle\nonumber\\
&&+|a_2\rangle|b_2\rangle)+\sqrt{|\delta|^2-|\gamma|^2}|a_3\rangle|b_2\rangle].
\end{eqnarray}
That is, Alice splits the parameters of the hyperentangled
Bell-class state in the spatial-mode DOF with UBS$_1$. One can see
that the spatial-mode state of the two-photon system $AB$ becomes a
maximally entangled one if the photon $A$ does not emit from the
spatial mode $a_3$.

Second, Alice transforms the polarization-mode state of the
two-photon system $AB$ into a maximally entangled one by splitting
the parameters of the  hyperentangled Bell-class state in the
polarization DOF, if the photon $A$ does not emit from the spatial
mode $a_3$. As shown in Fig. \ref{figure1}(a), Alice performs the
same polarization unitary operations on the two spatial modes $a_1$
and $a_2$. The  wave plate $R_\theta$ is used to rotate the
horizontal polarization $\vert H\rangle$ with an angle
$\theta=arccos(\beta/\alpha)$, that is, $\vert H\rangle\rightarrow
cos\theta \vert H\rangle +  sin\theta \vert V\rangle$. After the
photon $A$ coming from the two spatial modes $a_1$ and $a_2$ passes
through PBSs (i.e., PBS$_1$ and PBS$_2$) and $R_\theta$, the state
of the system is transformed from $|\varphi_1\rangle_{AB}$ into
$|\varphi_2\rangle_{AB}$. Here
\begin{eqnarray}                            
|\varphi_2\rangle_{AB}&=&[\beta(|H\rangle_A|H\rangle_B+|V\rangle_A|V\rangle_B)\nonumber\\
&&+\sqrt{|\alpha|^2-|\beta|^2}|V'\rangle_A|H\rangle_B]\nonumber\\
&&\otimes\gamma(|a_1\rangle|b_1\rangle+|a_2\rangle|b_2\rangle)\nonumber\\
&&+(\alpha|H\rangle_A|H\rangle_B+\beta|V\rangle_A|V\rangle_B)\nonumber\\
&&\otimes\sqrt{|\delta|^2-|\gamma|^2}|a_3\rangle|b_2\rangle.
\end{eqnarray}
Here $\vert V'\rangle$ presents the vertical polarization of the
photon after the operation  $R_\theta$.

Third, when the photon A passes through PBS$_3$ (PBS$_4$), DL, and
PBS$_5$ (PBS$_6$), the state of the two-photon system is transformed
from $|\varphi_2\rangle_{AB}$ into $|\varphi_3\rangle_{AB}$. Here
\begin{eqnarray}                            
|\varphi_3\rangle_{AB}&=&\beta\gamma(|H\rangle|H\rangle+|V\rangle|V\rangle)_{AB}(|a_1\rangle|b_1\rangle+|a_2\rangle|b_2\rangle)\nonumber\\
&&+\gamma\sqrt{|\alpha|^2-|\beta|^2}|V\rangle_A|H\rangle_B(|a'_1\rangle|b_1\rangle+|a'_2\rangle|b_2\rangle)\nonumber\\
&&+\sqrt{|\delta|^2-|\gamma|^2}(\alpha|H\rangle|H\rangle+\beta|V\rangle|V\rangle)_{AB}|a_3\rangle|b_2\rangle.\nonumber\\
\label{HBSoutcome}
\end{eqnarray}
From Eq. (\ref{HBSoutcome}), one can see that the state of the
two-photon system $AB$ becomes a maximally hyperentangled Bell state
$|\varphi_f\rangle_{AB}$ if the photon $A$ emits from the spatial
modes $a_1$ and $a_2$. Here
\begin{eqnarray}                           
|\varphi_f\rangle_{AB} =
\frac{1}{2}(|H\rangle|H\rangle+|V\rangle|V\rangle)_{AB}
(|a_1\rangle|b_1\rangle+|a_2\rangle|b_2\rangle).
\end{eqnarray}
If the photon $A$ emits from the spatial mode $a'_1$, $a'_2$, or
$a_3$, Alice and Bob cannot obtain a maximally hyperentangled Bell
state $|\varphi_f\rangle_{AB}$, which means this hyper-ECP fails. In
theory, Alice can judge whether this hyper-ECP succeeds or not,
according to the spatial mode of the photon $A$.

\begin{figure}[htb]                    
\centering
\includegraphics[width=7.2 cm]{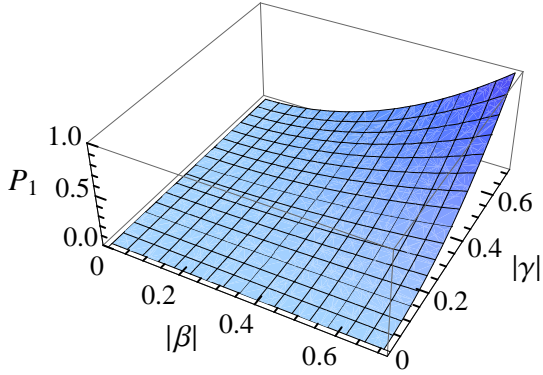}
\caption{(Color online) The success probability of our hyper-ECP for
each  two-photon four-qubit system in a partially hyperentangled
Bell-class state with known parameters.} \label{figure2}
\end{figure}

It is not difficult to calculate the success probability of  our
hyper-ECP for a hyperentangled Bell-class state. If the photon $A$
does not emit from the spatial mode $a'_1$,  $a'_2$, and $a_3$,  our
hyper-ECP succeeds, which takes place with the probability of
$P_1=4|\beta\gamma|^2$. Otherwise, the photon $A$ is detected by the
single-photon detectors and is destroyed, and our hyper-ECP fails.
That is, the total success probability of our hyper-ECP is
$P_1=4|\beta\gamma|^2$, shown in Fig. \ref{figure2}.

In a practical application of our hyper-ECP, there are two ways for
Alice and Bob, the two parties in quantum communication to judge
whether this hyper-ECP succeeds or not. On one hand, Alice can judge
that this hyper-ECP fails if one of the three detectors $D_1$,
$D_2$, and $D_3$ is clicked by the photon $A$. On the other hand, if
the efficiency of the single-photon detectors $D_i$ ($i=1,2,3$) is
100\%, Alice can also judge that this hyper-ECP succeeds if there
are no single-photon detectors clicked. At present, the efficiency
of a single-photon detector is lower than 100\%. That is, the case
that there is a photon which should be detected by the single-photon
detectors but not detected because of detection inefficiency, can be
mistaken as a successful event if Alice judges whether this
hyper-ECP succeeds or not with only the single-photon detectors
shown in Fig. \ref{figure1}. Fortunately, this mistaken case can be
eliminated by postselection, as the same as the entanglement
purification protocols
\cite{EPP3,EPPsimon,EPPexperiment,EPPsheng2,EPPdeng1} and the
entanglement concentration protocols \cite{ECP4,ECP5,ECP5a,ECP5b} in
only one DOF with linear-optical elements.  That is, this hyper-ECP
succeeds if Alice detects the photon  $A$  emitting from either the
spatial modes $a_1$ or $a_2$, when Alice and Bob use the photon pair
$AB$ to complete their task in quantum communication. Although the
photon pair $AB$ are detected in this time, the task of quantum
communication is also accomplished.

\begin{figure}[!h]
\centering\includegraphics[width=8 cm,angle=0]{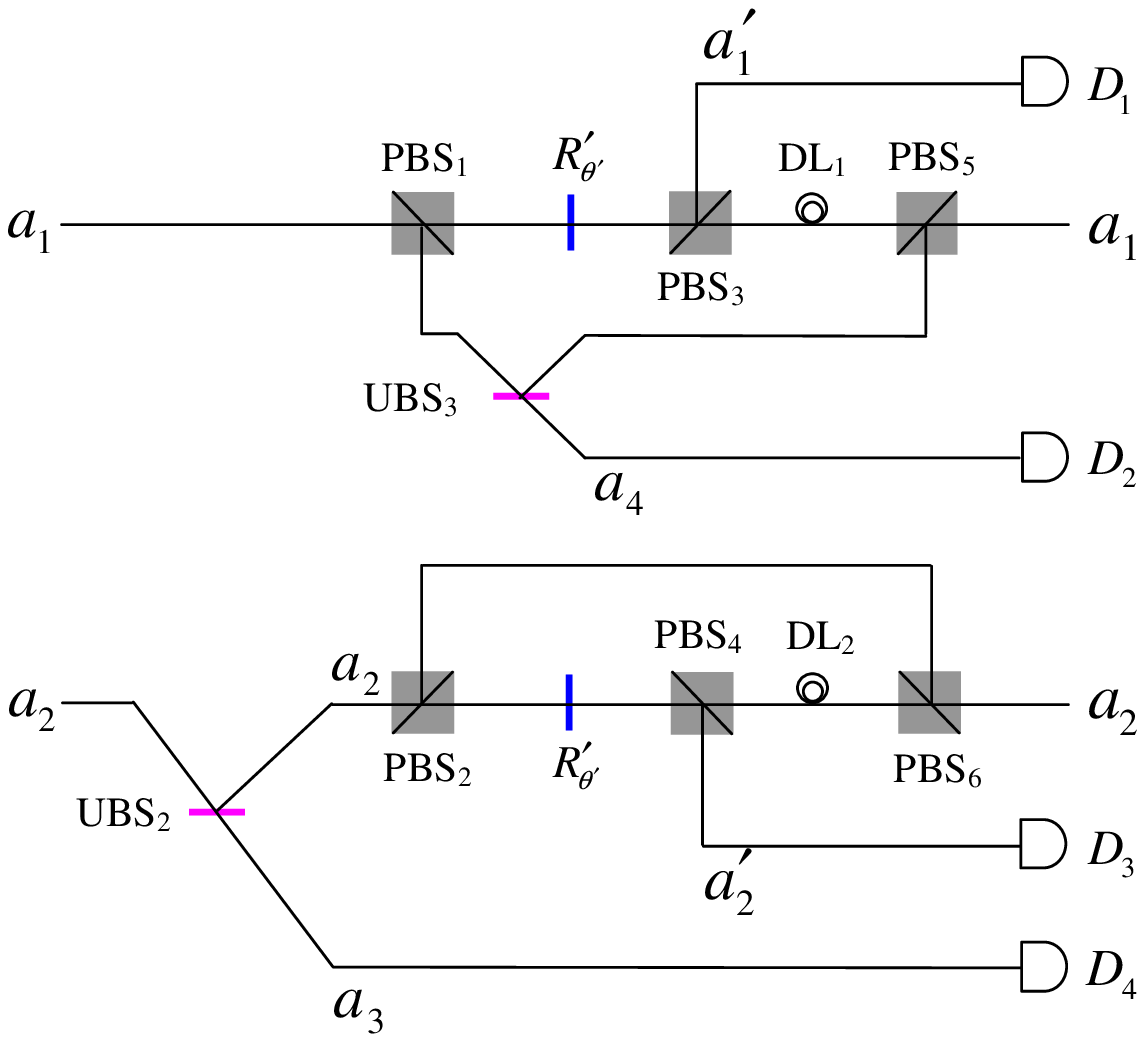}
\caption{(Color online) Schematic diagram of our hyper-ECP for
partially-entangled two-photon four-qubit cluster-class states with
known parameters. UBS$_2$ and UBS$_3$ represent two  unbalanced beam
splitters with the reflection coefficients $R_2=\alpha/\gamma$ and
$R_3=(\alpha\delta)/(\gamma\beta)$, respectively. $R_{\theta'}'$
represents a wave plate which can rotate the horizontal polarization
with an angle $\theta'=arccos(\delta/\gamma)$. DL$_1$ (DL$_2$)
denotes a time-delay device which is used to make the two
wavepackets reach the last PBS (PBS$_5$ or PBS$_6$) in each
Mach-Zehnder interferometer simultaneously. } \label{figure3}
\end{figure}

\subsection{Hyper-ECP for two-photon four-qubit systems in a partially hyperentangled cluster  state}
\label{sec22}

Suppose that the initial two-photon four-qubit partially
hyperentangled cluster-class state shared by two remote users Alice
and Bob is
\begin{eqnarray}                           
|\psi_0\rangle_{AB}&=&\alpha|H\rangle_A|H\rangle_B|a_1\rangle|b_1\rangle+\beta|V\rangle_A|V\rangle_B|a_1\rangle|b_1\rangle\nonumber\\
&&+\gamma|H\rangle_A|H\rangle_B|a_2\rangle|b_2\rangle-\delta|V\rangle_A|V\rangle_B|a_2\rangle|b_2\rangle.\nonumber\\
\end{eqnarray}
The four parameters $\alpha$, $\beta$, $\gamma$, and $\delta$ are
known to Alice and Bob, and they satisfy the relation
\begin{eqnarray}                           
|\alpha|^2+|\beta|^2+|\gamma|^2+|\delta|^2=1.
\end{eqnarray}
We assume $|\gamma| > |\alpha| > |\beta| > |\delta|$ below for
describing the principle of our hyper-ECP explicitly.

The principle of our hyper-ECP for a partially hyperentangled
two-photon four-qubit cluster-class state is shown in Fig.
\ref{figure3}, which is also implemented with linear-optical
elements only, by splitting the parameters of the cluster-class
state $|\psi_0\rangle_{AB}$. This hyper-ECP can also be accomplished
with local unitary operations on the photon $A$ in both the
spatial-mode and the polarization DOFs, and it includes three steps,
as shown in Fig. \ref{figure3}.

In the first step, only a spatial-mode unitary operation is
performed on the photon $A$ from the spatial mode $a_2 $ by using an
unbalanced BS (UBS$_2$) with the reflection coefficient
$R_2=\alpha/\gamma$. The partially-entangled two-photon four-qubit
cluster state $|\psi_0\rangle_{AB}$ is transformed into
$|\psi_1\rangle_{AB}$ in this step. Here
\begin{eqnarray}                             
|\psi_0\rangle_{AB}&=&\alpha|a_1\rangle|b_1\rangle(|H\rangle_A|H\rangle_B+\frac{\beta}{\alpha}|V\rangle_A|V\rangle_B)\nonumber\\
&&+\gamma|a_2\rangle|b_2\rangle(|H\rangle_A|H\rangle_B-\frac{\delta}{\gamma}|V\rangle_A|V\rangle_B),\nonumber\\
|\psi_1\rangle_{AB}&=&\alpha[|a_1\rangle|b_1\rangle(|H\rangle_A|H\rangle_B+\frac{\beta}{\alpha}|V\rangle_A|V\rangle_B)\nonumber\\
&&+|a_2\rangle|b_2\rangle(|H\rangle_A|H\rangle_B-\frac{\delta}{\gamma}|V\rangle_A|V\rangle_B)]\nonumber\\
&&+\sqrt{|\gamma|^2-|\alpha|^2}|a_3\rangle|b_2\rangle(|H\rangle_A|H\rangle_B\nonumber\\
&&-\frac{\delta}{\gamma}|V\rangle_A|V\rangle_B).
\end{eqnarray}

In the second step, Alice first performs a unitary operation with
PBS$_1$ on the photon $A$ emitting from the spatial mode $a_1$ and
then Alice performs a unitary operation on the photon $A$ coming
from the vertical polarization path of the spatial mode $a_1$, by
using an unbalanced BS (UBS$_3$) with the reflection coefficient
$R_3=(\alpha\delta)/(\gamma\beta)$. These two operations transform
the partially-entangled state $|\psi_1\rangle_{AB}$ into
$|\psi_2\rangle_{AB}$. Here,
\begin{eqnarray}                           
|\psi_2\rangle_{AB}&=&\alpha[|H\rangle_A|H\rangle_B(|a_1\rangle|b_1\rangle+|a_2\rangle|b_2\rangle)\nonumber\\
&&+\frac{\delta}{\gamma}|V\rangle_A|V\rangle_B(|a_1\rangle|b_1\rangle-|a_2\rangle|b_2\rangle]\nonumber\\
&&+ \alpha \sqrt{|\frac{\beta}{\alpha}|^2-|\frac{\delta}{\gamma}|^2} \;|V\rangle_A|V\rangle_B|a_4\rangle|b_1\rangle\nonumber\\
&&+\sqrt{|\gamma|^2-|\alpha|^2}\;|a_3\rangle|b_2\rangle(|H\rangle_A|H\rangle_B\nonumber\\
&&-\frac{\delta}{\gamma}|V\rangle_A|V\rangle_B).
\end{eqnarray}

The last step is used to perform unitary operations on the photon
$A$ in the polarization modes.  As shown in Fig. \ref{figure3}, the
polarization unitary operations performed on the photon $A$ by Alice
are the same ones for the two spatial modes $a_1$ and $a_2$, in
which $R'_{\theta'}$ contributes a rotation angle
$\theta'=arccos(\delta/\gamma)$ on the horizontal polarization of
the photon $A$. After photon $A$ passes through $R'_{\theta'}$ and
PBS, the state $|\psi_2\rangle_{AB}$ is changed to be
$|\psi_3\rangle_{AB}$. Here
\begin{eqnarray}                           
|\psi_3\rangle_{AB}&=&\frac{\alpha\delta}{\gamma}[|H\rangle_A|H\rangle_B(|a_1\rangle|b_1\rangle+|a_2\rangle|b_2\rangle)\nonumber\\
&&+|V\rangle_A|V\rangle_B(|a_1\rangle|b_1\rangle-|a_2\rangle|b_2\rangle]\nonumber\\
&&+\alpha\sqrt{1-|\frac{\delta}{\gamma}|^2}\;|V\rangle_A|H\rangle_B(|a'_1\rangle|b_1\rangle+|a'_2\rangle|b_2\rangle)\nonumber\\
&&+ \alpha \sqrt{|\frac{\beta}{\alpha}|^2-|\frac{\delta}{\gamma}|^2} \;|V\rangle_A|V\rangle_B|a_4\rangle|b_1\rangle\nonumber\\
&&+\sqrt{|\gamma|^2-|\alpha|^2}\;|a_3\rangle|b_2\rangle(|H\rangle_A|H\rangle_B\nonumber\\
&&-\frac{\delta}{\gamma}|V\rangle_A|V\rangle_B).\label{HCSoutcome}
\end{eqnarray}

From Eq. (\ref{HCSoutcome}), one can see that the state of the
two-photon system $AB$ becomes a maximally hyperentangled cluster
state $|\psi_f\rangle_{AB}$ if the photon $A$ emits from the spatial
modes $a_1$ and $a_2$. Here
\begin{eqnarray}                          
|\psi_f\rangle_{AB}&=&\frac{1}{2}[|H\rangle_A|H\rangle_B(|a_1\rangle|b_1\rangle+|a_2\rangle|b_2\rangle)\nonumber\\
&&+|V\rangle_A|V\rangle_B(|a_1\rangle|b_1\rangle-|a_2\rangle|b_2\rangle)].
\end{eqnarray}
If the photon $A$ emits from the spatial mode $a'_1$, $a'_2$, $a_3$,
or $a_4$, Alice and Bob cannot obtain a maximally hyperentangled
cluster state  $|\psi_f\rangle_{AB}$, which means this hyper-ECP
fails. In theory, Alice can judge whether this hyper-ECP succeeds or
not, according to the spatial modes of the photon $A$. As the same
as that in our hyper-ECP for a partially hyperentangled Bell state,
Alice and Bob can also judge whether this hyper-ECP succeeds or
fails by postselection if the efficiencies of the single-photon
detectors are low than 100\% in a practical application.

\begin{figure}[htb]                    
\centering
\includegraphics[width=7.2 cm]{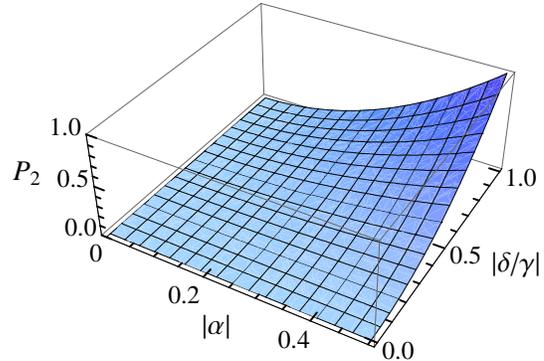}
\caption{(Color online) The success probability of our hyper-ECP for
each two-photon four-qubit system in a partially hyperentangled
cluster-class state with known parameters.} \label{figure4}
\end{figure}

It is not difficult to calculate the success probability of  our
hyper-ECP for a hyper-entangled two-photon cluster-class state.  If
one of the four detectors  $D_1$, $D_2$, $D_3$, and $D_4$ clicks, at
least one DOF of the partially entangled state is projected into a
product state, which means that this hyper-ECP  fails.  If none of
the four detectors  $D_1$, $D_2$, $D_3$, and $D_4$  clicks in
principle, the partially entangled cluster-class state
$|\psi_0\rangle_{AB}$ is transformed into the maximally entangled
two-photon four-qubit cluster state $|\psi_f\rangle_{AB}$  with a
success probability of $P_2=4|\alpha\delta/\gamma|^2$, shown in Fig.
\ref{figure4}.

%

\begin{figure}[!h]
\centering
\includegraphics[width=7.8 cm,angle=0]{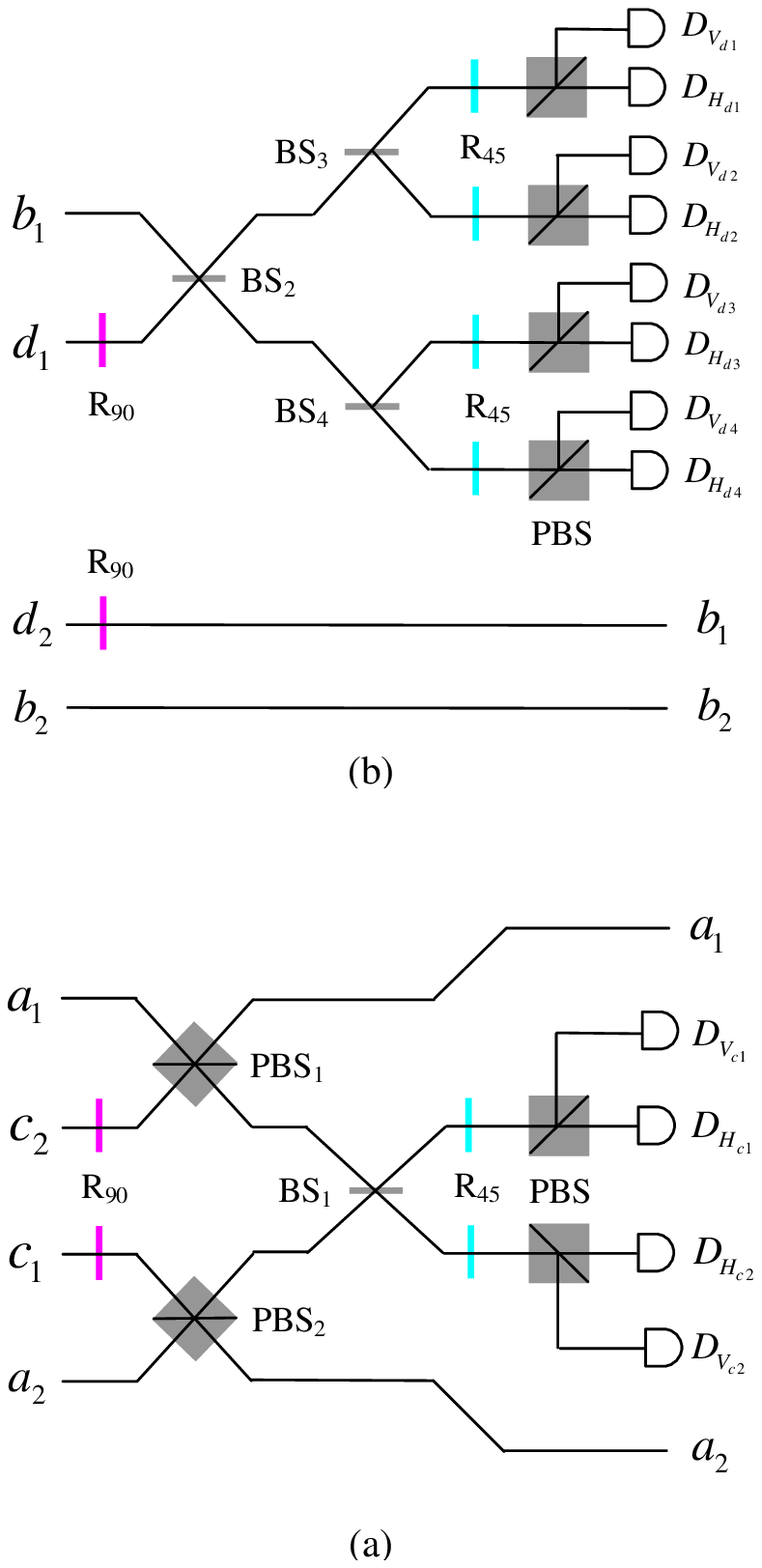}
\caption{(Color online) Schematic diagram of our hyper-ECP for
partially hyperentangled Bell states with unknown parameters. (a)
Operations performed by Alice. (b) Operations performed by Bob.
R$_{90}$ represents a half-wave plate which is used to perform a
polarization bit-flip operation $X=|H\rangle\langle V| +
|V\rangle\langle H|$. R$_{45}$ represents a half-wave plate which is
used to perform a  Hadamard operation on the polarization DOF of
photons.} \label{figure5}
\end{figure}

\section{ Hyper-ECP for two-photon four-qubit systems in an  unknown hyperentangled pure state }
\label{sec3}

In this section, we will discuss the hyper-ECPs for
partially-entangled two-photon four-qubit pure states in both the
spatial-mode and the polarization DOFs with their parameters unknown
to the two remote users Alice and Bob, including partially
hyperentangled Bell states and cluster states. In this time, the
parameter-splitting method does not work as the parameters are
unknown to Alice and Bob. However, these hyper-ECPs can be
accomplished with the Schmidt projection method, although they are
more complex than those \cite{ECP1,ECP4,ECP5} for the photon systems
in only partially entangled polarization states.

\subsection{Hyper-ECP for two-photon four-qubit systems in an unknown hyperentangled Bell-class state}
\label{sec31}

Suppose that there are two identical two-photon systems $AB$ and
$CD$ in a partially hyperentangled Bell-class state
\begin{eqnarray}                           
|\phi_0\rangle_{AB}&=&(\alpha|H\rangle_A|H\rangle_B+\beta|V\rangle_A|V\rangle_B)\nonumber\\
&&\otimes(\gamma|a_1\rangle|b_1\rangle+\delta|a_2\rangle|b_2\rangle),\nonumber\\
|\phi_0\rangle_{CD}&=&(\alpha|H\rangle_C|H\rangle_D+\beta|V\rangle_C|V\rangle_D)\nonumber\\
&&\otimes(\gamma|c_1\rangle|d_1\rangle+\delta|c_2\rangle|d_2\rangle).
\end{eqnarray}
Here  subscripts $A$ and $B$ represent the two photons in a
partially hyperentangled state $|\phi_0\rangle_{AB}$, and the
subscripts $C$ and $D$ represent another  two photons in the
partially hyperentangled state $|\phi_0\rangle_{CD}$. $i_1$ and
$i_2$ are the two spatial modes of the photon $i$ ($i=A,B,C,D$). The
photons $A$ and $C$ belong to Alice, and the photons $B$ and $D$
belong to Bob. The four parameters $\alpha$, $\beta$, $\gamma$, and
$\delta$ are unknown to Alice and Bob, and they satisfy the relation
$|\alpha|^2+|\beta|^2=|\gamma|^2+|\delta|^2=1$.

The principle of our hyper-ECP for partially hyperentangled
Bell-class states with unknown parameters is shown in Fig.
\ref{figure5}. The initial state of the four-photon system $ABCD$
can be rewritten as follows:
\begin{eqnarray}                          
|\Phi_0\rangle&=&|\phi_0\rangle_{AB}\otimes|\phi_0\rangle_{CD}\nonumber\\
&=&(\alpha^2|H\rangle_A|H\rangle_B|H\rangle_C|H\rangle_D\nonumber\\
&&+\alpha\beta|V\rangle_A|V\rangle_B|H\rangle_C|H\rangle_D\nonumber\\
&& +\alpha\beta|H\rangle_A|H\rangle_B|V\rangle_C|V\rangle_D\nonumber\\
&&+\beta^2|V\rangle_A|V\rangle_B|V\rangle_C|V\rangle_D)\nonumber\\
&&  \otimes(\gamma^2|a_1\rangle|b_1\rangle|c_1\rangle|d_1\rangle+\gamma\delta|a_2\rangle|b_2\rangle|c_1\rangle|d_1\rangle\nonumber\\
&&+\gamma\delta|a_1\rangle|b_1\rangle|c_2\rangle|d_2\rangle+\delta^2|a_2\rangle|b_2\rangle|c_2\rangle|d_2\rangle).\;\;\;\;\;\;
\end{eqnarray}
After Alice and Bob flip the polarizations of the photons $C$ and
$D$ on both the two spatial modes with half wave plates ($R_{90}$),
the state of the four-photon system becomes
\begin{eqnarray}                          
|\Phi_1\rangle&=& (\alpha^2|H\rangle_A|H\rangle_B|V\rangle_C|V\rangle_D\nonumber\\
&& + \alpha\beta|V\rangle_A|V\rangle_B|V\rangle_C|V\rangle_D\nonumber\\
&& + \alpha\beta|H\rangle_A|H\rangle_B|H\rangle_C|H\rangle_D\nonumber\\
&& + \beta^2|V\rangle_A|V\rangle_B|H\rangle_C|H\rangle_D)\nonumber\\
&& \otimes(\gamma^2|a_1\rangle|b_1\rangle|c_1\rangle|d_1\rangle+\gamma\delta|a_2\rangle|b_2\rangle|c_1\rangle|d_1\rangle\nonumber\\
&& +
\gamma\delta|a_1\rangle|b_1\rangle|c_2\rangle|d_2\rangle+\delta^2|a_2\rangle|b_2\rangle|c_2\rangle|d_2\rangle).\;\;\;\;\;\;
\label{eq14}
\end{eqnarray}
Subsequently,  Alice puts the photons from the spatial modes $a_1$
and  $c_2$  into PBS$_1$, and those emitting from  $a_2$ and $c_1$
into  PBS$_2$. Bob puts the photons emitting from the spatial modes
$b_1$ and $d_1$  into  BS$_2$.  Here PBSs in Fig. \ref{figure5}(a)
are used to complete a parity-check measurement on the polarization
DOF of the two photons, and BS in Fig.
 \ref{figure5}(b) is used to complete a parity-check measurement on
the spatial-mode DOF of the two photons with the HOM effect. In
detail, if the two photons $A$ and $C$ have the same polarizations
$\vert H\rangle_A\vert H\rangle_C$ or $\vert V\rangle_A\vert
V\rangle_C$ (called each of them an even-parity state in the
polarization DOF), there is one and only one photon which will be
detected by the single-photon detectors shown in Fig.
\ref{figure5}(a) in principle after the two photons pass through the
PBSs (PBS$_1$ and PBS$_2$). Otherwise, there are none or two photons
which will be detected by the detectors in principle if the two
photons have different polarization states (called them the
odd-parity polarization states). If the two photons $B$ and $D$ have
different spatial-mode states (the odd-parity states), there is one
and only one photon which will be detected by the single-photon
detectors shown in Fig. \ref{figure5}(b) in principle.

These parity-check measurements by PBSs and BSs divide the states of
the four-photon system into two groups, based on both the spatial
mode and the polarization DOFs. Alice and Bob pick up the
even-parity terms of the polarization DOF of the photon pair $AC$ in
Alice's hand with the same parameter and the odd-parity terms of the
spatial-mode DOF of the photon pair $CD$ in Bob's hand with the same
parameter (these instances will lead the fact that both Alice and
Bob will have only one detector clicked). That is, the state of the
four-photon system with the selected terms becomes
\begin{eqnarray}                           
|\Phi_2\rangle&=&\frac{1}{2}(|V\rangle_A|V\rangle_B|V\rangle_C|V\rangle_D+|H\rangle_A|H\rangle_B|H\rangle_C|H\rangle_D)\nonumber\\
&&\otimes
(|a_2\rangle|b_2\rangle|c_1\rangle|d_1\rangle+|a_1\rangle|b_1\rangle|c_2\rangle|d_2\rangle).
\label{eq15}
\end{eqnarray}
In the other cases, this hyper-ECP fails.

At last, both Alice and Bob perform  Hadamard operations on the
spatial-mode and the polarization DOFs of the photons $C$ and $D$
with a half-wave plate $R_{45}$ and a BS, respectively. The selected
terms shown in Eq. (\ref{eq15}) are transformed into
\begin{eqnarray}                           
|\Phi_3\rangle&=&\frac{1}{4}[(|V\rangle|V\rangle+|H\rangle|H\rangle)_{AB}(|V\rangle|V\rangle+|H\rangle|H\rangle)_{CD}\nonumber\\
&&+(|H\rangle|H\rangle-|V\rangle|V\rangle)_{AB}(|H\rangle|V\rangle+|V\rangle|H\rangle)_{CD}]\nonumber\\
&&\otimes[(|a_2\rangle|b_2\rangle+|a_1\rangle|b_1\rangle)(|c_1\rangle|d_1\rangle+|c_2\rangle|d_2\rangle)\nonumber\\
&&-(|a_1\rangle|b_1\rangle-|a_2\rangle|b_2\rangle)(|c_1\rangle|d_2\rangle+|c_2\rangle|d_1\rangle)].\nonumber\\
\end{eqnarray}
If the outcomes of the two clicked detectors are in the even-parity
polarization modes and the even-parity spatial modes, the state of
photon pair $AB$ is projected into the maximally Bell hyperentangled
state
$|\varphi_f\rangle_{AB}=\frac{1}{2}(|H\rangle_A|H\rangle_B+|V\rangle_A|V\rangle_B)(|a_1\rangle|b_1\rangle+|a_2\rangle|b_2\rangle)$.
If it is an odd-parity outcome in the measurement on the photon pair
$CD$ in the polarization (spatial-mode) DOF, a phase-flip operation
$\sigma^p_z=|H\rangle\langle H| - |V\rangle\langle V|$
($\sigma^s_z=|b_1\rangle\langle b_1| - |b_2\rangle\langle b_2|$) on
the photon $B$ is required to obtain the state
$|\varphi_f\rangle_{AB}$.

\begin{figure}[htb]                    
\centering
\includegraphics[width=8 cm]{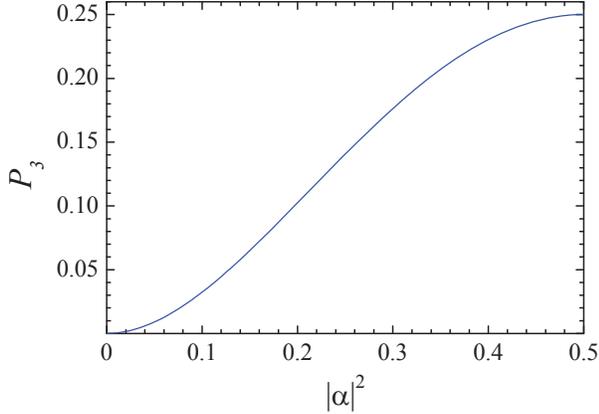}
\caption{(Color online) Success probability of our hyper-ECP for a
pair of two-photon systems in a partially hyperentangled Bell state
with unknown parameters. The relations of the parameters of the
partially hyperentangled Bell state are chosen as
$|\alpha|=|\gamma|$ and $|\beta|=|\delta|$.} \label{figure6}
\end{figure}

In principle, if both Alice and Bob have only one detector clicked,
the hyper-ECP   succeeds  with the probability
$P_3=4|\alpha\beta\gamma\delta|^2$ (shown in Fig. \ref{figure6} for
the cases with $|\alpha|=|\gamma|$ and $|\beta|=|\delta|$).
Otherwise, the hyper-ECP fails. In a practical application of this
hyper-ECP, Alice and Bob can also judge whether this hyper-ECP
succeeds or fails by postselection if the efficiencies of the
single-photon detectors are not 100\%.

\subsection{Hyper-ECP for two-photon four-qubit systems in an unknown hyperentangled cluster-class state}
\label{sec32}

We assume that there are two identical two-photon four-qubit systems
in the partially Bell-type hyperentangled cluster-class states,
\begin{eqnarray}                            
|\Psi_0\rangle_{AB}&=&\alpha|H\rangle_A|H\rangle_B(\gamma|a_1\rangle|b_1\rangle+\delta|a_2\rangle|b_2\rangle)\nonumber\\
&&+\beta|V\rangle_A|V\rangle_B(\gamma|a_1\rangle|b_1\rangle-\delta|a_2\rangle|b_2\rangle),\nonumber\\
|\Psi_0\rangle_{CD}&=&\alpha|H\rangle_C|H\rangle_D(\gamma|c_1\rangle|d_1\rangle+\delta|c_2\rangle|d_2\rangle)\nonumber\\
&&+\beta|V\rangle_C|V\rangle_D(\gamma|c_1\rangle|d_1\rangle-\delta|c_2\rangle|d_2\rangle).\;\;\;\;\;\;\label{eq21}
\end{eqnarray}
Here the subscripts $A$, $B$, $C$, and $D$ represent four photons.
The two photons $A$ and $C$ belong to Alice, and the two photons $B$
and $D$ belong to Bob. The four parameters $\alpha$, $\beta$,
$\gamma$, and $\delta$ are unknown to Alice and Bob, and they
satisfy the relation,
\begin{eqnarray}                           
|\alpha|^2+|\beta|^2=|\gamma|^2+|\delta|^2=1.
\end{eqnarray}

\begin{figure}[!h]
\centering
\includegraphics[width=6 cm,angle=0]{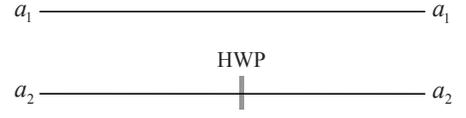}
\caption{Schematic diagram of the operation for transforming a
maximally hyperentangled Bell state into a maximally hyperentangled
two-photon cluster state. HWP represents a half-wave plate which is
used to perform a polarization phase-flip operation
$U_p=|H\rangle\langle H|-|V\rangle\langle V|$ on the photon when it
emits from the spatial mode $a_2$.} \label{figure7}
\end{figure}

It is suitable to implement this hyper-ECP for an unknown partially
hyperentangled two-photon four-qubit cluster-class state with the
quantum circuit shown in Fig.  \ref{figure5}. Alice and Bob perform
the polarization parity-check measurement and the spatial-mode
parity-check measurement on the two photon pairs $AC$ and $BD$,
respectively. The even-parity terms of the polarization DOF of the
photon pair in Alice's hand and the odd-parity terms of the
spatial-mode DOF of the photon pair in Bob's hand are picked up.
After the detections and the local phase-flip operations, the state
of the two-photon system $AB$ becomes $
|\varphi_f\rangle=\frac{1}{2}(|H\rangle_A|H\rangle_B+|V\rangle_A|V\rangle_B)
\otimes (|a_1\rangle|b_1\rangle+|a_2\rangle|b_2\rangle)$ which is a
maximally hyperentangled Bell state. This maximally hyperentangled
Bell state can be transformed into a nonlocal maximally entangled
two-photon four-qubit cluster state $
|\psi_f\rangle_{AB}=\frac{1}{2}(|H\rangle_A|H\rangle_B|a_1\rangle|b_1\rangle+|V\rangle_A|V\rangle_B|a_1\rangle|b_1\rangle
+|H\rangle_A|H\rangle_B|a_2\rangle|b_2\rangle-|V\rangle_A|V\rangle_B|a_2\rangle|b_2\rangle)$
with the operation shown in Fig. \ref{figure7}.

Up to now, we have discussed the hyper-ECP for a partially
hyperentangled Bell-type two-photon four-qubit cluster-class state
and its success probability is $P_4=4|\alpha\beta\gamma\delta|^2$
with only linear-optical elements and single-photon detectors,  the
same as that in the hyper-ECP for a hyperentangled Bell-class state
with unknown parameters.  The hyperentanglement concentration of
two-photon four-qubit systems in an arbitrary unknown partially
hyperentangled cluster-class state is more complex than that in a
partially hyperentangled Bell-type two-photon four-qubit
cluster-class state, as discussed in   Appendix \ref{appendixa}. In
a practical application, Alice and Bob can first transmit a
hyperentangled Bell state over a quantum channel and concentrate it
with the hyper-ECP discussed in Sec. \ref{sec31}, and then they
transform a nonlocal maximally hyperentangled Bell state into a
nonlocal maximally hyperentangled two-photon four-qubit cluster
state. This modification can reduce the resource needed.

\section{Discussion and summary}
\label{sec4}

We have proposed four hyper-ECPs for two classes of two-photon
four-qubit states partially entangled in both the spatial-mode and
the polarization DOFs of two-photon systems in two kinds of
conditions with linear-optical elements. The two classes of
two-photon states are hyperentangled Bell states and cluster states,
respectively, and the two kinds of conditions discussed are the
two-photon states with known parameters and those with unknown
parameters, respectively.

In the first kind of condition about the two-photon four-qubit
states with their parameters known to the two remote legitimate
users, only one of the two parties is required to perform local
operations in our hyper-ECPs. Here the unbalanced BS and the half
wave plates for rotating the horizontal polarization are required to
decrease the parameters of the terms with large probabilities until
they are equal to the parameters of the terms with the smallest
probability. We call this  the parameter-splitting method for
entanglement concentration. With single-photon detectors, the
parties in quantum communication can read out whether the hyper-ECPs
succeed or not in principle. If none of the single-photon detectors
clicks, the two hyper-ECPs succeed. The success probabilities of
these two hyper-ECPs are shown in Fig.
 \ref{figure2} and Fig.  \ref{figure4}, respectively. It shows that
the success  probability of the hyper-ECP for a partially
hyperentangled Bell state is determined by the two smallest
parameters in the two DOFs, and the success probability of the
hyper-ECP for a partially entangled two-photon cluster state is
correlated to the relation of three parameters of the partially
hyperentangled state.

In the second kind of condition about the two-photon four-qubit
states with their parameters unknown to the two legitimate users,
both the users are required to perform some quantum operations on
their photons, and two copies of the unknown partially entangled
states are required. In these two hyper-ECPs, the polarization
parity-check measurement and the spatial-mode parity-check
measurement are performed with some PBSs and BSs, respectively. In
these two hyper-ECPs, two copies of photon pairs in the unknown
state are required for each round of concentration, and the two
legitimate users can make their parity-check measurements
synchronously. For the concentration of a partially entangled
two-photon four-qubit cluster state, an auxiliary step is required
to transform a  nonlocal  maximally hyperentangled Bell state into a
nonlocal  maximally entangled two-photon cluster state with a
linear-optical element. These two hyper-ECPs succeed when both Alice
and Bob have one single-photon detector clicked, and local
phase-flip operations are needed if the outcomes of the two
single-photon  detectors are not in even-parity polarization modes
and even-parity spatial modes. We have discussed the success
probabilities of these two hyper-ECPs for some special states shown
in Fig. \ref{figure6}. It shows that the success probabilities of
the two hyper-ECPs change with the parameters of the partially
hyperentangled states.

In a practical quantum communication, Alice and Bob can obtain the
information about the parameters of a nonlocal partially
hyperentangled pure state  if they measure a sufficient number of
sample photon pairs. It is not difficult to find that the two
hyper-ECPs with our parameter-splitting method are more efficient
and practical than those with the Schmidt projection method if there
are large numbers of quantum data needed to be exchanged between the
two parties in quantum communication. First, the former can be used
to concentrate each of the two-photon four-qubit systems, while the
latter require two copies of the two-photon four-qubit systems in
the same partially entangled pure state. Second, the efficiency of
the former is more than two times  that of the latter. Third, the
quantum circuits in the former are far simpler than those in the
latter. Contrarily, if there are a small quantity of  quantum data
needed to be exchanged between the two parties, the hyper-ECPs with
our parameter-splitting method as they do not require the two
parties to measure the samples for obtaining the accurate
information about the parameters of the partially hyperentangled
pure state.

On one hand, although we have introduced our parameter-splitting
method for the hyperentanglement concentration of two-photon
four-qubit systems in nonlocal partially hyperentangled pure states
with known parameters, it is obviously suitable for all the
entanglement concentration on photon systems in nonlocal partially
entangled pure states with known parameters, including those with
one DOF and those with several DOFs, no matter what the form of the
state is and what the number of the particles in the system is. The
similar tasks can be accomplished with linear optics only. On the
other hand, this way may be suitable for atom systems, electron-spin
systems, and others if the parties in quantum communication can
construct the elements similar to the UBSs and PBSs in photon
systems. It is a general way for efficient entanglement
concentration. If we use our parameter-splitting method for the
concentration of photonic polarization entanglement, it is far
simpler than the ECPs existing for photon systems in a
less-entangled polarization state with known parameters, including
those based on entanglement swapping \cite{ECP2}, a collective
unitary evolution \cite{ECP3}, or additional single photons
\cite{ECP7,ECP8,ECP11}.

In our four hyper-ECPs, we assume the efficiency of the
linear-optical elements,  such as PBSs, BSs,  wave plates, and
half-wave plates, to be perfect. That is, there is no photon loss in
these linear-optical elements. Moreover, the efficiency of the
single-photon detectors is assumed to be 100\%.  In a practical
application of these hyper-ECPs, they do not work in an ideal
condition. The nonideal elements and detectors will decrease the
success probabilities of these hyper-ECPs. In this time, the two
parties can obtain maximally hyperentangled states by postselection
yet. In detail, by picking up the cases in which there is a photon
emitting from each of the output ports $a_1$ ($a_2$) and $b_1$
($b_2$), the two parties can make their hyper-ECP work in a
practical condition,  as the same as the entanglement purification
protocols \cite{EPP3,EPPsimon,EPPexperiment,EPPsheng2,EPPdeng1} and
the entanglement concentration protocols
\cite{ECP4,ECP5,ECP5a,ECP5b} in only one DOF with linear optical
elements.

In summary, we have proposed four hyper-ECPs for two classes of
two-photon four-qubit states entangled in two DOFs in two kinds of
conditions. With these four hyper-ECPs, the  parties in quantum
communication can concentrate nonlocal partially hyperentangled Bell
states and nonlocal partially entangled two-photon cluster states
with their parameters known or unknown to the two remote users. As
hyperentangled states can increase the channel capacity of
long-distance quantum communication processing, our hyper-ECPs for
two-photon states entangled in two DOFs may be very useful in
long-distance quantum communication in the future.

The task of entanglement concentration is to distill  some quantum
systems in a  nonlocal  maximally entangled state from those in a
nonlocal  partially entangled pure state for two remote users
\cite{ECP1}. The goal of this work is focused on the nonlocal
hyperentanglement concentration of two-photon four-qubit systems in
a partially hyperentangled pure state. However, if the two-photon
four-qubit systems are in a mixed hyperentangled state, it is more
complex for the two remote users Alice and Bob to obtain some
quantum systems in a hyperentangled state with a higher fidelity
than the original one, as discussed in  Appendix \ref{appendixb}. Of
course, this is the goal of entanglement purification \cite{EPP1}.
For a general hyperentanglement purification, nonlinearity is
required \cite{HEPP}. As a simple example of the parameter-splitting
method for the entanglement concentration of photon  systems in one
DOF, we discuss the concentration for polarization entangled states
and spatial-mode entangled states independently in  Appendix
\ref{appendixc}.

\section*{ACKNOWLEDGMENTS}

This work is supported by the National Natural Science Foundation of
China under Grant No. 11174039 and  NECT-11-0031.

\appendix

\section{Hyper-ECP for two-photon four-qubit systems in an
arbitrary unknown hyperentangled cluster-class state}
\label{appendixa}

We assume that there are four identical two-photon systems in an
arbitrary partially hyperentangled cluster state,
\begin{eqnarray}                           
|\Psi_0\rangle_{AB}&=&\alpha|H\rangle_A|H\rangle_B|a_1\rangle|b_1\rangle+\beta|V\rangle_A|V\rangle_B|a_1\rangle|b_1\rangle\nonumber\\
&+&\gamma|H\rangle_A|H\rangle_B|a_2\rangle|b_2\rangle-\delta|V\rangle_A|V\rangle_B|a_2\rangle|b_2\rangle,\nonumber\\
|\Psi_0\rangle_{CD}&=&\alpha|H\rangle_C|H\rangle_D|c_1\rangle|d_1\rangle+\beta|V\rangle_C|V\rangle_D|c_1\rangle|d_1\rangle\nonumber\\
&+&\gamma|H\rangle_C|H\rangle_D|c_2\rangle|d_2\rangle-\delta|V\rangle_C|V\rangle_D|c_2\rangle|d_2\rangle,\nonumber\\
|\Psi_0\rangle_{A'B'}&=&\alpha|H\rangle_{A'}|H\rangle_{B'}|a'_1\rangle|b'_1\rangle+\beta|V\rangle_{A'}|V\rangle_{B'}|a'_1\rangle|b'_1\rangle\nonumber\\
&+&\gamma|H\rangle_{A'}|H\rangle_{B'}|a'_2\rangle|b'_2\rangle-\delta|V\rangle_{A'}|V\rangle_{B'}|a'_2\rangle|b'_2\rangle,\nonumber\\
|\Psi_0\rangle_{C'D'}&=&\alpha|H\rangle_{C'}|H\rangle_{D'}|c'_1\rangle|d'_1\rangle+\beta|V\rangle_{C'}|V\rangle_{D'}|c'_1\rangle|d'_1\rangle\nonumber\\
&+&\gamma|H\rangle_{C'}|H\rangle_{D'}|c'_2\rangle|d'_2\rangle-\delta|V\rangle_{C'}|V\rangle_{D'}|c'_2\rangle|d'_2\rangle.\nonumber\\
\end{eqnarray}
Here the subscripts $AB$, $CD$, $A'B'$, and $C'D'$ represent four
photon pairs.  The four photons $A$, $C$, $A'$, and $C'$ belong to
Alice, and the four photons $B$, $D$, $B'$, and $D'$ belong to Bob.
The four parameters $\alpha$, $\beta$, $\gamma$, and $\delta$ are
unknown to Alice and Bob, and they satisfy the relation
$|\alpha|^2+|\beta|^2+|\gamma|^2+|\delta|^2=1$.

The principle of our hyper-ECP for an arbitrary  partially
hyperentangled two-photon four-qubit cluster state  with unknown
parameters includes three steps.  Let us describe them in detail
below.

First, Alice and Bob divide the four photon pairs into two groups,
that is, $AB$ and  $CD$ in one group, and  $A'B'$   and $C'D'$ in
the other group. Alice and Bob perform the same operations on these
two groups. The quantum circuit shown in Fig.  \ref{figure5} without
all the half-wave plates $R_{90}$  is used to select the terms with
the same parameter, resorting to its parity-check effect. The
even-parity terms of the polarization DOF of photon pairs in Alice's
hand and the odd-parity terms of the spatial-mode DOF of photon
pairs in Bob's hand are picked up. After the detections on $CD$
($C'D'$) and the local phase-flip operations on the photon $B$
($B'$), the states of the photon pairs $AB$ and $A'B'$ become
\begin{eqnarray}                           
|\Psi_1\rangle_{AB}&=&\frac{|\alpha\gamma|}{\sqrt{2(|\alpha\gamma|^2+|\beta\delta|^2)}}(|a_1\rangle|b_1\rangle+|a_2\rangle|b_2\rangle)\nonumber\\
&&\otimes(|H\rangle_A|H\rangle_B-\frac{\beta\delta}{\alpha\gamma}|V\rangle_A|V\rangle_B),\nonumber\\
|\Psi_1\rangle_{A'B'}&=&\frac{|\alpha\gamma|}{\sqrt{2(|\alpha\gamma|^2+|\beta\delta|^2)}}(|a'_1\rangle|b'_1\rangle+|a'_2\rangle|b'_2\rangle)\nonumber\\
&&\otimes(|H\rangle_{A'}|H\rangle_{B'}-\frac{\beta\delta}{\alpha\gamma}|V\rangle_{A'}|V\rangle_{B'}).
\end{eqnarray}

Now, the spatial modes of the two-photon systems are in a maximally
entangled state, and the next step is to let the polarization modes
become a  maximally entangled one. This task can be completed with
the quantum circuit shown in Fig.  \ref{figure5}.  The polarization
parity-check measurement is performed on the photon pair $AA'$ in
Alice's hand, and the spatial-mode parity-check measurement is
performed on the photon pair $BB'$ in Bob's hand. Then the
even-parity terms of the polarization DOF of photon pairs in Alice's
hand and the odd-parity terms of the spatial-mode DOF of photon
pairs in Bob's hand are picked up. After the detections on $A'B'$
and the local phase-flip operations on the photon $B$, the state of
the photon pair $AB$ becomes
$|\varphi_f\rangle_{AB}=\frac{1}{2}(|H\rangle_A|H\rangle_B+|V\rangle_A|V\rangle_B)(|a_1\rangle|b_1\rangle+|a_2\rangle|b_2\rangle)$
which is a maximally hyperentangled Bell state. This maximally
hyperentangled Bell state can be transformed into a maximally
entangled two-photon cluster state
$|\psi_f\rangle_{AB}=\frac{1}{2}(|H\rangle_A|H\rangle_B|a_1\rangle|b_1\rangle
+|V\rangle_A|V\rangle_B|a_1\rangle|b_1\rangle+|H\rangle_A|H\rangle_B|a_2\rangle|b_2\rangle-|V\rangle_A|V\rangle_B|a_2\rangle|b_2\rangle)$
with the operation shown in Fig. \ref{figure7}.

Up to now, we have transformed an arbitrary partially entangled
two-photon cluster state into a maximally hyperentangled two-photon
cluster state with the success  probability
$P_4=\frac{4|\alpha\beta\gamma\delta|^2}{2(|\alpha\gamma|^2+|\delta\beta|^2)}$
by exploiting some linear-optical elements only. The principle of
the first two steps of our hyper-ECP for partially hyperentangled
two-photon cluster states is the same as that in the hyper-ECP for
partially hyperentangled Bell states, and an arbitrary partially
entangled two-photon cluster state can be changed to be a maximally
hyperentangled Bell one. In the third step, the maximally
hyperentangled Bell state is transformed into a maximally entangled
two-photon cluster state with a linear-optical element.

In this hyper-ECP, one can see that four photon pairs are required
for concentrating a hyperentangled two-photon four-qubit
cluster-class state with four arbitrary unknown parameters. On one
hand, it is not efficient, compared with the hyper-ECP for a
partially hyperentangled Bell state as the latter requires only two
pairs in a practical application. Moreover, this is, in principle,
the necessary condition for the hyperentanglement concentration of
an arbitrary unknown two-photon four-qubit cluster state because
there are three independent parameters which are unknown to Alice
and Bob, while there are only two independent parameters which are
unknown to Alice and Bob in the hyperentanglement concentration of
an unknown partially hyperentangled Bell state. On the other hand,
it is possible to prepare a nonlocal maximally hyperentangled
cluster-class state with a nonlocal maximally hyperentangled Bell
state using a linear-optical element shown in Fig. \ref{figure7}.
Therefore, in a practical application in quantum communication, the
hyper-ECP for the hyperentangled Bell state is sufficient for
long-distance quantum communication with hyperentangled two-photon
four-qubit states. After the parties obtain a nonlocal maximally
hyperentangled Bell state, they need only transform it into a
nonlocal maximally hyperentangled cluster state with linear-optical
elements.

\section{Entanglement purification for a spatial-polarization mixed hyperentangled Bell state}
\label{appendixb}

We assume that there are two identical two-photon four-qubit systems
in a mixed hyperentangled state,
\begin{eqnarray}                           
\rho_{AB}&=&[F_1|\phi^+_{AB}\rangle_P\langle\phi^+_{AB}|+(1-F_1)|\psi^+_{AB}\rangle_P\langle\psi^+_{AB}|]\nonumber\\
&&\otimes|\phi^+_{AB}\rangle_S\langle\phi^+_{AB}|,\nonumber\\
\rho_{CD}&=&[F_1|\phi^+_{CD}\rangle_P\langle\phi^+_{CD}|+(1-F_1)|\psi^+_{CD}\rangle_P\langle\psi^+_{CD}|]\;\;\;\;\;\;\nonumber\\
&&\otimes|\phi^+_{CD}\rangle_S\langle\phi^+_{CD}|.
\end{eqnarray}
Here the subscripts $AB$ and $CD$ represent two photon pairs. The
two photons $A$ and $C$ belong to Alice, and the two photons $B$ and
$D$ belong to Bob. The subscripts $P$ and $S$ represent the
polarization and the spatial-mode DOFs, respectively.  The four
states $|\phi^+\rangle_P$, $|\psi^+\rangle_P$, $|\phi^+\rangle_S$,
and $|\psi^+\rangle_S$ are defined as
\begin{eqnarray}                            
|\phi^+\rangle_P=\frac{1}{\sqrt{2}}(|HH\rangle+|VV\rangle),\nonumber\\
|\psi^+\rangle_P=\frac{1}{\sqrt{2}}(|HV\rangle+|VH\rangle),\nonumber\\
|\phi^+\rangle_S=\frac{1}{\sqrt{2}}(|a_1b_1\rangle+|a_2b_2\rangle),\nonumber\\
|\psi^+\rangle_S=\frac{1}{\sqrt{2}}(|a_1b_2\rangle+|a_2b_1\rangle).
\end{eqnarray}
The state of the four-photon system $ABCD$ can be described as
\begin{eqnarray}                            
\rho&=&\rho_{AB}\otimes\rho_{CD}.
\end{eqnarray}
This mixed hyperentangled Bell state can be viewed as the mixture of
four  maximally hyperentangled Bell states, the same as the
conventional entanglement purification protocols for photon pairs
\cite{EPP1,EPP2,EPP3,EPPsimon,EPPexperiment,EPPsheng1,EPPsheng2,EPPsheng3,EPPdeng1}.
That is, it is the mixture of the four pure states in the
polarization DOF:
$|\phi^+_{AB}\rangle_P\otimes|\phi^+_{CD}\rangle_P$ with the
probability $F_1^2$,
$|\phi^+_{AB}\rangle_P\otimes|\psi^+_{CD}\rangle_P$ with the
probability $F_1(1-F_1)$,
$|\psi^+_{AB}\rangle_P\otimes|\phi^+_{CD}\rangle_P$ with the
probability $F_1(1-F_1)$, and
$|\psi^+_{AB}\rangle_P\otimes|\psi^+_{CD}\rangle_P$ with the
probability $(1-F_1)^2$. The state of the system in the spatial-mode
DOF is $|\phi^+_{AB}\rangle_S\otimes|\phi^+_{CD}\rangle_S$.

\begin{figure}[!h]
\centering
\includegraphics[width=6.6 cm,angle=0]{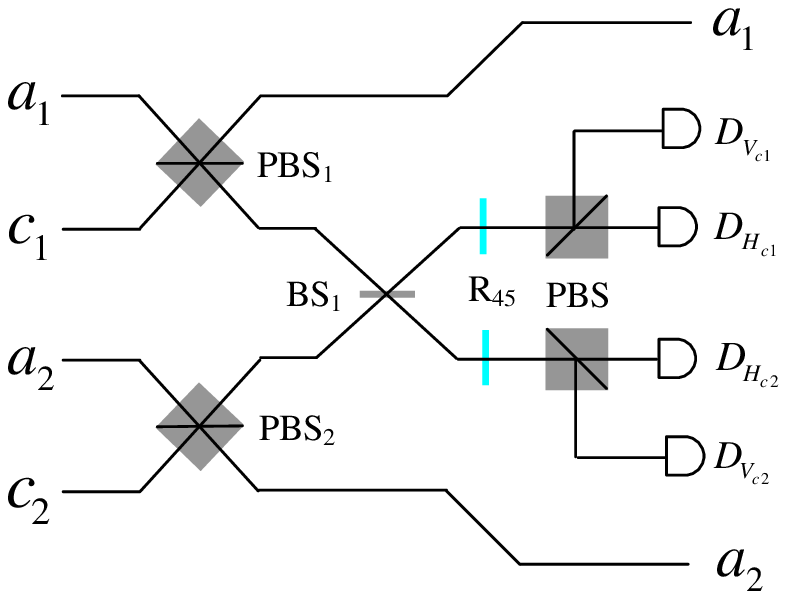}
\caption{(Color online)  Schematic diagram of our hyperentanglement
purification protocol (hyper-EPP) for a mixed hyperentangled Bell
state with polarization bit-flip errors. Bob performs the same
operations as Alice by replacing the photons $A$ and $C$ with the
photons $B$ and $D$, respectively. } \label{figure8}
\end{figure}

In order to implement this hyper-EPP,  Alice and Bob have to perform
the same operations on the two photon pairs $AC$ and $BD$, shown in
Fig. \ref{figure8}. That is, Alice (Bob)  performs the polarization
parity-check measurements and the spatial-mode parity-check
measurements on the  photon pair $AC$ ($BD$) first and then detects
the  photon  $C$ ($D$). If both Alice and Bob detect one and only
one photon with their single-photon detectors,  the state of the
two-photon system $AB$ becomes
\begin{eqnarray}                           
\rho_f&=&[F'_1|\phi^+\rangle_P\langle\phi^+|+(1-F'_1)|\psi^+\rangle_P\langle\psi^+|]_{AB}\nonumber\\
&&\otimes|\phi^+\rangle_S\langle\phi^+|_{AB}.
\end{eqnarray}
Here $F'_1=F_1^2/[F_1^2+(1-F_1)^2]$. If $F_1>1/2$, $F'_1>F_1$.  That
is, the fidelity of the maximally hyperentangled Bell state
$|\varphi_f\rangle_{AB}=\frac{1}{2}(|H\rangle_A|H\rangle_B+|V\rangle_A|V\rangle_B)(|a_1\rangle|b_1\rangle+|a_2\rangle|b_2\rangle)$
increases after a round of entanglement purification if  $F_1>1/2$.
In a practical application, Alice and Bob should resort to
postselection for judging whether there is one and only one photon
detected by each of them or not, similar to the conventional EPPs in
the polarization DOF
\cite{EPP1,EPP2,EPP3,EPPsimon,EPPexperiment,EPPsheng1}.

Our hyper-EPP is used to purify the mixed hyperentangled Bell state
with polarization bit-flip errors. The polarization phase-flip error
can be transformed into the polarization bit-flip error with local
operations \cite{EPP1,EPP2,EPP3,EPPsimon,EPPexperiment,EPPsheng1}.
The spatial-mode entanglement is robust against bit-flip errors
because it is hard for a photon to permeate into another fiber
\cite{EPPsimon,EPPexperiment}. The phase-flip error of the
spatial-mode DOF can be depressed with current technology
\cite{EPPsimon,EPPexperiment}. If there are both the polarization
bit-flip error and the spatial-mode bit-flip error in a mixed
hyperentangled Bell state, the present hyper-EPP cannot be
implemented with linear optics, which is limited by the ability of
linear optics for purifying only one qubit error. In this time, we
should use nonlinear optics to implement the hyper-EPP for a mixed
hyperentangled Bell state with both the polarization bit-flip error
and the spatial-mode bit-flip error \cite{HEPP}.

\section{Parameter-splitting-based entanglement concentration of photon systems in a known partially entangled state in one DOF}
\label{appendixc}

In this section, we will describe the ECPs for a partially entangled
polarization Bell state and a partially entangled spatial-mode Bell
state independently based on our parameter-splitting method in
detail.

\begin{figure}[htbp]             
\centering\includegraphics[width=6.4 cm]{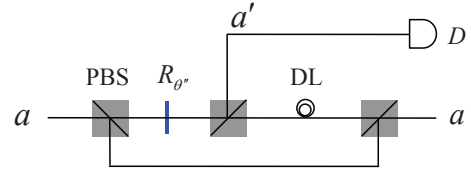} \caption{(Color
online) Schematic diagram of our parameter-splitting-based ECP for a
partially entangled Bell state with known parameters in the
polarization DOF. $R_{\theta''}$ represents a wave plate which can
rotate the horizontal polarization with an angle
$\theta''=arccos(\beta/\alpha)$. $D$ represents a single-photon
detector.} \label{figure9}
\end{figure}

\subsection{Parameter-splitting-based entanglement concentration of photon systems in a known partially entangled polarization state}
\label{appendixc1}

Let us assume that the initial partially entangled Bell-class state
in the polarization DOF is
\begin{eqnarray}                            
|\varphi_0\rangle_{AB}=\alpha|H\rangle_A|H\rangle_B+\beta|V\rangle_A|V\rangle_B.
\end{eqnarray}
Here the subscripts $A$ and $B$ represent the two photons shared by
the two remoter users Alice and Bob, respectively. $\alpha$ and
$\beta$ are two real parameters that are known to Alice and Bob, and
they satisfy the relation $|\alpha|^2+|\beta|^2=1$ and
$|\alpha|>|\beta|$.

The principle of the ECP for a partially entangled polarization Bell
state is shown in Fig. \ref{figure9}, which can be implemented by
performing local unitary operations on the photon $A$ in its
polarization DOF. Alice performs the polarization unitary operation
$R_{\theta''}$ on the photon $A$, which rotates the horizontal
polarization $\vert H\rangle$ with an angle
$\theta''=arccos(\beta/\alpha)$, that is, $\vert H\rangle\rightarrow
cos\theta'' \vert H\rangle +  sin\theta'' \vert V\rangle$. After the
photon $A$ passes through PBS and $R_{\theta''}$, the state of the
system is transformed from $|\varphi_0\rangle_{AB}$ into
$|\varphi_1\rangle_{AB}$. Here
\begin{eqnarray}                            
|\varphi_1\rangle_{AB}&=&\frac{1}{\sqrt{2}}[\beta(|H\rangle_A|H\rangle_B+|V\rangle_A|V\rangle_B)\nonumber\\
&&+\sqrt{|\alpha|^2-|\beta|^2}|V'\rangle_A|H\rangle_B].
\end{eqnarray}
Here $\vert V'\rangle$ presents the vertical polarization of the
photon after the operation  $R_{\theta''}$, and the photon will emit
from the spatial mode $a'$ and it will be detected by the
single-photon detector $D$ in principle. If the detector $D$ does
not click, the state $|\varphi_1\rangle_{AB}$ becomes a maximally
entangled one. That is, the state of the two-photon system $AB$
becomes
\begin{eqnarray}                           
|\phi^+\rangle_{AB} =
\frac{1}{\sqrt{2}}(|H\rangle|H\rangle+|V\rangle|V\rangle)_{AB}.
\end{eqnarray}
This is just the maximally entangled polarization Bell state of a
two-photon system. If the detector $D$ clicks, the polarization
state $|\varphi_1\rangle_{AB}$ is projected into the product state
$|V'\rangle_A|H\rangle_B$. Certainly, the photon $A$ is destroyed,
and the entanglement concentration fails in this time.

Now we have implemented our parameter-splitting-based ECP for a
partially entangled polarization Bell state  with known parameters.
If the detector $D$ does not click, our ECP succeeds, which takes
place with the probability of $P_p=2|\beta|^2$. If the detector $D$
clicks, the photon $A$ is destroyed and our ECP fails.

\begin{figure}[htbp]             
\centering\includegraphics[width=5.5 cm]{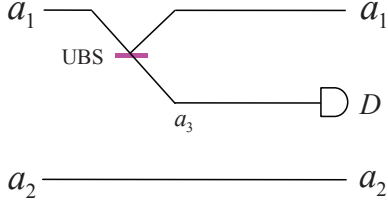}
\caption{(Color
online) Schematic diagram of our parameter-splitting-based ECP for a
partially entangled Bell state with known parameters in the
spatial-mode DOF. UBS represents an unbalanced beam splitter with
the reflection coefficient $R=\beta/\alpha$.} \label{figure10}
\end{figure}

\subsection{Parameter-splitting-based entanglement concentration of photon systems in a known partially entangled spatial-mode state}
\label{appendixc2}

Let us assume that the initial partially entangled Bell-class state
in the spatial-mode DOF is
\begin{eqnarray}                            
|\varphi_0\rangle_{AB}=\alpha|a_1\rangle|b_1\rangle+\beta|a_2\rangle|b_2\rangle.
\end{eqnarray}
Here the subscripts $A$ and $B$ represent the two photons shared  by
Alice and Bob. $\alpha$ and $\beta$ are two real parameters that are
known to Alice and Bob, and they satisfy the relation
$|\alpha|^2+|\beta|^2=1$ and $|\alpha|>|\beta|$.

The principle of the ECP for a partially entangled spatial-mode Bell
state is shown in Fig. \ref{figure10}.  It can be implemented by
performing local unitary operations on the photon $A$ in the
spatial-mode DOF. Alice performs a unitary operation on the spatial
mode $a_1$ by using an unbalanced BS \cite{UBS} with the reflection
coefficient $R=\beta/\alpha$, shown in Fig. \ref{figure1}(b), and
the partially entangled spatial-mode Bell state
$|\varphi_0\rangle_{AB}$ is changed to be $|\varphi_1\rangle_{AB}$.
Here
\begin{eqnarray}                             
|\varphi_1\rangle_{AB}&=&\beta(|a_1\rangle|b_1\rangle
+|a_2\rangle|b_2\rangle)\nonumber\\
&&+\sqrt{|\alpha|^2-|\beta|^2}|a_3\rangle|b_1\rangle.
\end{eqnarray}
The state of the two-photon system $AB$  becomes a maximally
entangled one if the detector $D$ does not click. That is, the state
of the two-photon system $AB$ becomes
\begin{eqnarray}                            
|\varphi_2\rangle_{AB}=\frac{1}{\sqrt{2}}(|a_1\rangle|b_1\rangle+|a_2\rangle|b_2\rangle).
\end{eqnarray}
If the detector $D$ clicks, the state of the system is projected
into a product state $|a_3\rangle|b_2\rangle$. In this time, the
photon $A$ is destroyed, and the entanglement concentration fails.

It is not difficult to calculate the success probability of our ECP
for a partially entangled spatial-mode Bell state. If the detector
$D$ does not click, our ECP succeeds, which takes place with the
probability of $P_s=2|\beta|^2$. If the detector $D$ clicks, the
photon $A$ is destroyed and our ECP fails.

\end{document}